\documentclass{jfm}

\usepackage{graphicx}
\usepackage{natbib}

\ifCUPmtlplainloaded \else
  \checkfont{eurm10}
  \iffontfound
    \IfFileExists{upmath.sty}
      {\typeout{^^JFound AMS Euler Roman fonts on the system,
                   using the 'upmath' package.^^J}%
       \usepackage{upmath}}
      {\typeout{^^JFound AMS Euler Roman fonts on the system, but you
                   dont seem to have the}%
       \typeout{'upmath' package installed. JFM.cls can take advantage
                 of these fonts,^^Jif you use 'upmath' package.^^J}%
      }
  \else
  \fi
\fi

\ifCUPmtlplainloaded \else
  \checkfont{msam10}
  \iffontfound
    \IfFileExists{amssymb.sty}
      {\typeout{^^JFound AMS Symbol fonts on the system, using the
                'amssymb' package.^^J}%
       \usepackage{amssymb}%
       \let\le=\leqslant  
       \let\ge=\geqslant  
      }{}
  \fi
\fi

\ifCUPmtlplainloaded \else
  \IfFileExists{amsbsy.sty}
    {\typeout{^^JFound the 'amsbsy' package on the system, using it.^^J}%
     \usepackage{amsbsy}}
    {\providecommand\boldsymbol[1]{\mbox{\boldmath $##1$}}}
\fi

\newsavebox{\astrutbox}
\sbox{\astrutbox}{\rule[-5pt]{0pt}{20pt}}

\title[Compressible Air Entrapment in High-Speed Drop Impacts on Solid Surfaces]{Compressible Air Entrapment in High-Speed Drop Impacts on Solid Surfaces}

\author[Yuan LIU, Peng TAN \and Lei XU]
{Y\ls U\ls A\ls N\ns L\ls I\ls U,\ns P\ls E\ls N\ls G\ns T\ls A\ls N\ns \and L\ls E\ls I\ns X\ls U\ls\thanks{Email address for correspondence: xulei@phy.cuhk.edu.hk}\ns}

\affiliation{Department of physics, The Chinese University of Hong Kong, Shatin, Hong Kong, China}
\pubyear{2010}
\volume{650}
\pagerange{119--126}
\date{?; revised ?; accepted ?. - To be entered by editorial office}

\begin{document}

\maketitle

\begin{abstract}
Using high-speed photography coupled with optical interference, we
experimentally study the air entrapment during a liquid drop
impacting a solid substrate. We observe the formation of a
compressed air film before the liquid touches the substrate, with
internal pressure considerably higher than the atmospheric value.
The degree of compression highly depends on the impact velocity, as
explained by balancing the liquid deceleration with the large
pressure of compressed air. After contact, the air film expands
vertically at the edge, reducing its pressure within a few tens of
microseconds and producing a thick rim on the perimeter. This
thick-rimmed air film subsequently contracts into an air bubble,
governed by the complex interaction between surface tension, inertia
and viscous drag. Such a process is universally observed for impacts
above a few centimeters high.
\end{abstract}

\section{Introduction}
The impacts of liquid drops onto solid substrates are ubiquitous and
appear in a variety of applications, such as spray coating and ice
accumulation on aircraft \citep{Yarin2006ARFM}. While the
interaction between liquid and solid has been extensively studied,
the important role of air is discovered only recently, as shown by
the surprising finding that air pressure strongly influences the
liquid splash outcomes \citep{Xu2005PRL, Xu2007PRE}. Therefore,
understanding the behavior of air during liquid-solid impacts will
bring new advancement to this fundamental phenomenon, and may
benefit practical processes such as splash control and surface
coating. Experimentally, entrapment of air during impacts is
commonly observed: when the impact speed is high ($\sim1m/s$), a
thin air film is trapped at the very beginning, which subsequently
contracts into one or two air bubbles \citep{Thoroddsen2003JFM,
Thoroddsen2005JFM}. At low impact speed ($\sim0.1m/s$), however, the
air under the drop remains connected with outside for the majority
of time (i.e., not entirely trapped), until it gets enclosed by the
moving contact line \citep{Jolet2012PRL}. Recent study
\citep{John2012PRL} further reveals a short-lived nanometer thick
air film right before the formation of contact line. The air
entrapment depends on the impact velocity, the liquid property
\citep{Thoroddsen2003JFM, Thoroddsen2005JFM, Jolet2012PRL,
John2012PRL} and the surrounding air pressure \citep{Sid2011PRL}.
Theoretically, simulations have explored air entrapment with both
compressible \citep{Brenner2009PRL, Brenner2010JFM} and
incompressible \citep{Smith2003JFM, Peter2010JFM} models. Marked
deformation of drop surface \emph{before} it touches the solid is
predicted: the drop surface is deformed upward at the impact center,
making the first contact away from the center on a ring-like area.
Compressible model by \cite{Brenner2009PRL} and
\cite{Brenner2010JFM} further indicates significant compression in
the trapped air, resulting in a pressure considerably higher than
the atmospheric value. Despite these important predictions, however,
experimental measurement on the exact condition of trapped air is
still missing, especially near the critical moment of impact. In
particular, even the fundamental question whether the trapped air is
compressed or not remains unclearly. To clarify these puzzles and
better understand the impact phenomenon, experimental study on air
entrapment close to the moment of impact is highly desirable.

\section{Experimental methods}\label{sec:exp}
Using fast photography coupled with optical interference
\citep{Sid2011PRL, Jolet2012PRL}, we experimentally study the air
entrapment during a liquid drop impacting a smooth substrate at
relatively high speeds ($0.7 - 3\mathrm{m/s}$). We observe the
formation of a compressed air film \emph{before} the liquid touches
the substrate, with internal pressure considerably higher than the
atmospheric value. The degree of compression highly depends on the
impact velocity, which is explained by balancing the liquid
deceleration with the large pressure of compressed air. After
contact, the air film expands vertically at the edge, reducing its
pressure within a few tens of microseconds and producing a thick rim
on the perimeter. This thick-rimmed air film subsequently contracts
into an air bubble, governed by the complex interaction between
surface tension, inertia and viscous effects (see Fig.5(a) for the
entire process). Such a process is universally observed for impacts
above a few centimeters high.

To independently study the effects of viscosity and surface tension,
three different liquids are used: $\mathrm{H_2O}$, oil-1.04 and
oil-9.30. The two oils are silicone oils with similar surface
tensions $\sigma$ but different dynamic viscosities:
$\mathrm{\mu=1.04 mPa\cdot s}$ and $\mathrm{9.30 mPa\cdot s}$. By
contrast, $\mathrm{H_2O}$ and oil-1.04 have similar viscosities but
different surface tensions (see table \ref{tab:dl}). Reproducible
drops around millimeter in size (see table \ref{tab:dl} for details)
are released from rest at different heights, and impact a smooth and
dry cover glass at various velocities ($\mathrm{0.7 - 3m/s}$). The
two silicon oils completely wet the glass substrate with zero static
contact angle, and $\mathrm{H_2O}$ has the static contact angle
$\theta = 65^\circ\pm5^\circ$. All experiments are performed at the
atmospheric pressure, $\mathrm{P_0=101kPa}$. The impacts are viewed
from below with an inverted microscope and recorded with a
high-speed camera (Photron SA4) at recording speeds up to 150,000
frames per second. The illumination light is monochromatic, with the
wavelength $\mathrm{\lambda=546nm}$ and coherence length a few
microns. The short coherence length makes sure that there is no
interference between the two sides of the substrate. The Newton's
rings produced by the trapped air can quantitatively characterize
the thickness profile of air.

\begin{table}
  \begin{center}
\def~{\hphantom{0}}
  \begin{tabular}{lccccc}
      liquid & $\mathrm{\rho(kg/m^{3})}$ & $\mathrm{\sigma(mN/m)}$ & $\mathrm{\mu(mPa\cdot s)}$ & $\theta$(degree) & R(mm)\\
      $\mathrm{H_2O}$ & 1000 & $50\pm5$ & $1.00\pm0.01$ & $65\pm5$ & $2.0\pm0.1$ ($1.5\pm0.1$)\\
      oil-1.04 & 816 & 17.4 & $1.04\pm0.01$ & $0$ & $1.5\pm0.1$ ($1.2\pm0.1$)\\
      oil-9.30 & 930 & 20.1 & $9.30\pm0.02$ & $0$ & $1.5\pm0.1$\\
  \end{tabular}
  \caption{\label{tab:example}material properties and drop radius of different
liquids. R is measured right before contact by fitting the local
radius of curvature at the bottom of the drop. For the lowest
release height, strong oscillation in the drop makes $\mathrm{R}$
different from the average radius in $\mathrm{H_2O}$ and oil-1.04,
as shown in the parenthesis.}
  \label{tab:dl}
  \end{center}
\end{table}

\section{Experimental results}\label{sec:res_paper}
\subsection{Formation of air film before contact ($t\le0$)}

\begin{figure}
\begin{center}
\includegraphics[width=0.8\textwidth]{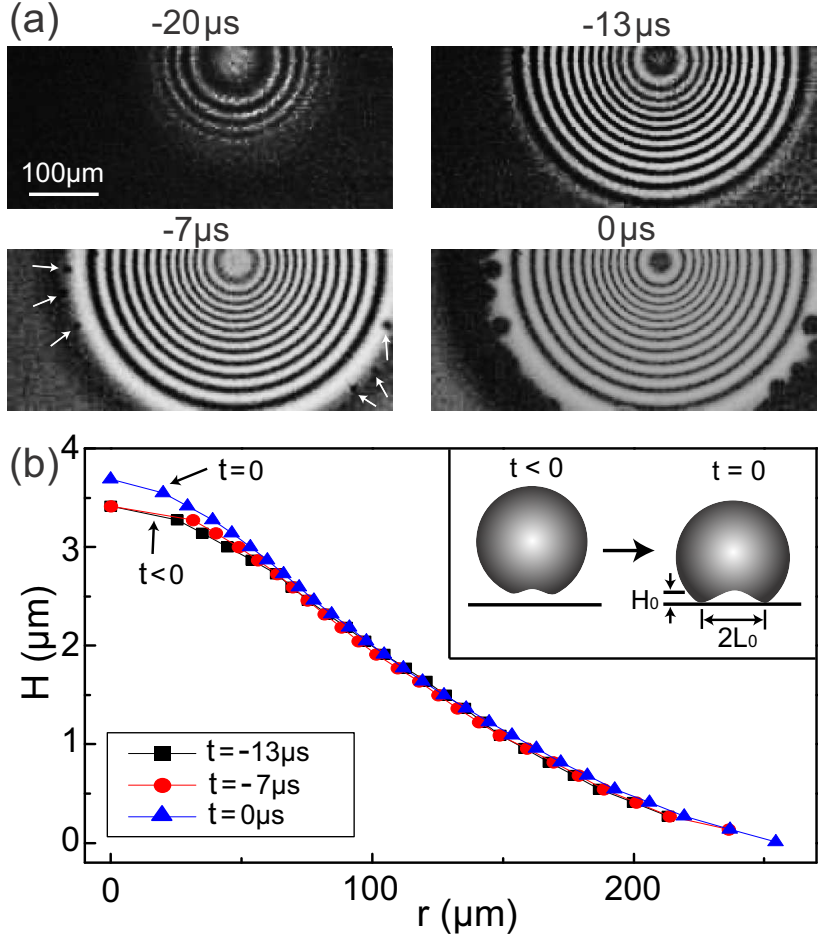}
\caption{(color online) The formation of trapped air film before
liquid touches solid. (\textit{a}) The interference patterns before
contact by an oil-9.30 drop at impact velocity
$\mathrm{V_0=0.74\pm0.01}$ m/s. The time uncertainty is half of the
frame interval, $4\mu s$. In the first frame ($\mathrm{-20\mu s}$),
Newton's rings appear as the drop-to-substrate distance becomes
smaller than the coherence length of light. In the second frame
($\mathrm{-13\mu s}$), a dimple with horizontal scale several
hundred microns but vertical scale a few microns is produced. The
third frame ($\mathrm{-7\mu s}$) reveals a few black spots of
isolated contacts at the edge (specified by the white arrows). The
fourth frame ($\mathrm{0\mu s}$) shows the global liquid-solid
contact on a thick black ring at the edge of the pattern. This
moment of first global contact is also defined as the moment of
$\mathrm{t=0}$ throughout our study. (\textit{b}) The thickness
profile, H(r), constructed from the patterns in (a). Note the very
different scales for x and y axes. The curves almost overlap with
each other, indicating very little motion of the interface through
different frames. Significant difference only appears at impact
center ($\mathrm{r=0}$), where the interface moves \emph{upward}
instead of downward, with the speed $\mathrm{0.02\pm0.01m/s}$. The
schematics in the inset illustrates a dimple with horizontal scale
$\mathrm{2L_0}$ and thickness $\mathrm{H_0}$ (drawing not on
scale).}
\end{center}
\end{figure}

We first clarify the entrapment behavior \emph{before} liquid
touches solid. As the drop approaches the substrate, the air in the
thin gap can not escape immediately and gets compressed to a high
pressure (exact values measured later in Fig.4(b)). Such a high
pressure locally pushes in the drop surface and creates a dimple
around impact center, as illustrated by the cartoon in Fig.1(b)
inset. The interference patterns from the dimple are shown in
Fig.1(a), for an oil-9.30 drop at the impact speed
$\mathrm{V_0=0.74\pm0.01 m/s}$. In the first frame ($\mathrm{-20\mu
s}$), Newton's rings appear as the drop-to-substrate distance
becomes smaller than the coherence length of light (a few microns).
In the second frame ($\mathrm{-13\mu s}$), a quite flat dimple with
the horizontal scale several hundred microns but the vertical scale
a few microns is produced. Because of this thin and flat nature, the
dimple can be exactly considered as an air film. The third frame
($\mathrm{-7\mu s}$) shows a very similar pattern as the second one,
except with a few black spots of isolated contacts indicated by the
white arrows. The fourth frame ($\mathrm{0\mu s}$) reveals the
global liquid-solid contact on a thick black ring at the edge of the
pattern. This moment of first global contact is also defined as the
moment of $\mathrm{t=0}$ throughout our study.

The thickness of the dimple, $\mathrm{H}$, can be accurately derived
from the pattern with the relationship $\mathrm{\Delta
H=\lambda/4=136.5nm}$ between two neighboring rings. Using contact
area as the zero-thickness reference point, we can quantitatively
determine the entire thickness profile of the trapped air in the
$\mathrm{t=0}$ frame. We can further determine the profiles for the
two previous frames ($\mathrm{t=-7\mu s}$ and $\mathrm{-13\mu s}$),
due to their highly similar patterns as the $\mathrm{t=0}$ frame,
which enables the tracking of every ring. The thickness as a
function of distance from the center, $\mathrm{H}$ versus
$\mathrm{r}$, is plotted in Fig.1(b): $\mathrm{13\mu s}$ before the
contact, an air film with the lateral size $\mathrm{2L_0\cong500\mu
m}$ but a thickness of only $\mathrm{H_0\cong3.5\mu m}$ forms, which
maintains an almost identical profile through the next several
frames. Significant variation only appears near the center
($\mathrm{r\cong0}$), where the interface moves \emph{upward}
instead of downward, with a small speed $\mathrm{0.02\pm0.01m/s}$
(calculated from the profiles at $\mathrm{t=-7\mu s}$ and
$\mathrm{t=0\mu s}$). Clearly, the motion of the drop is quite
complex immediately before the contact: while the main body falls at
the impact speed, $\mathrm{V_0=0.74m/s}$; the small volume above the
trapped air ($\mathrm{2L_0 \cong 500\mu m}$) stays almost
stationary, with the region at $\mathrm{r\cong0}$ even moving
oppositely in upward direction.

\subsection{Profile of air film upon contact ($t=0$)}

To systematically study the air film properties, we characterize
their profiles at $\mathrm{t=0}$ for different velocities and
liquids, as shown in Fig.2(a). The three panels correspond to the
three different liquids and each panel contains data for several
impact velocities. In general, air film becomes thinner with the
increase of impact velocity, as illustrated by the measurements. For
quantitative understanding, we plot the maximum thickness measured
at $\mathrm{r=0}$, $\mathrm{H_0}$, as a function of $\mathrm{V_0}$
in Fig.2(b). Clearly $\mathrm{H_0}$ decreases with $\mathrm{V_0}$,
consistent with the previous theory (\cite{Brenner2009PRL}).
However, the exact dependence of $\mathrm{H_0}$ on $\mathrm{V_0}$
differs from the prediction: we calculate the ``compressible
factor'' defined theoretically as
$\mathrm{\epsilon=P_0/(R\mu_g^{-1}V_0^7\rho^4)^{1/3}}$(\cite{Brenner2009PRL},
$\mu_g$ is the dynamic viscosity of air), and obtain the range
$\mathrm{\epsilon^{-1}\sim}$ 0.1--10 for our experimental data. Thus
our experiments are at the transition region from ``incompressible
regime'' to ``compressible regime'' predicted by the theory, and
$\mathrm{H_0}$ should decrease with $\mathrm{V_0}$ faster than the
power law of $\mathrm{H_0\sim V_0^{-2/3}}$. However the log-log
plots in Fig.2(b) find powers much slower than -2/3. This
discrepancy calls for further theoretical studies on the problem.

We also measure the horizontal radius, $\mathrm{L_0}$, versus
$\mathrm{V_0}$ for different liquids and compare it with the
existing theory (\cite{Peter2012JFM}). In Fig.2(c), we plot our
measurements as solid symbols and the theoretical prediction,
$\mathrm{L_0=3.8\cdot(4\mu_g/\rho V_0)^{1/3}R^{2/3}}$
(\cite{Peter2012JFM}), as the open symbols. We use the bottom radius
of curvature right before contact shown in Table 1 for the
calculation. Without any fitting parameter, a reasonable agreement
is observed at high $\mathrm{V_0}$ while some deviation occurs for
low $\mathrm{V_0}$. This is surprising since the theory is based on
an incompressible calculation, which should match our low
$\mathrm{V_0}$ rather than the high $\mathrm{V_0}$ region. Further
study is required to clarify this puzzle.

\begin{figure}
\begin{center}
\includegraphics[width=0.8\textwidth]{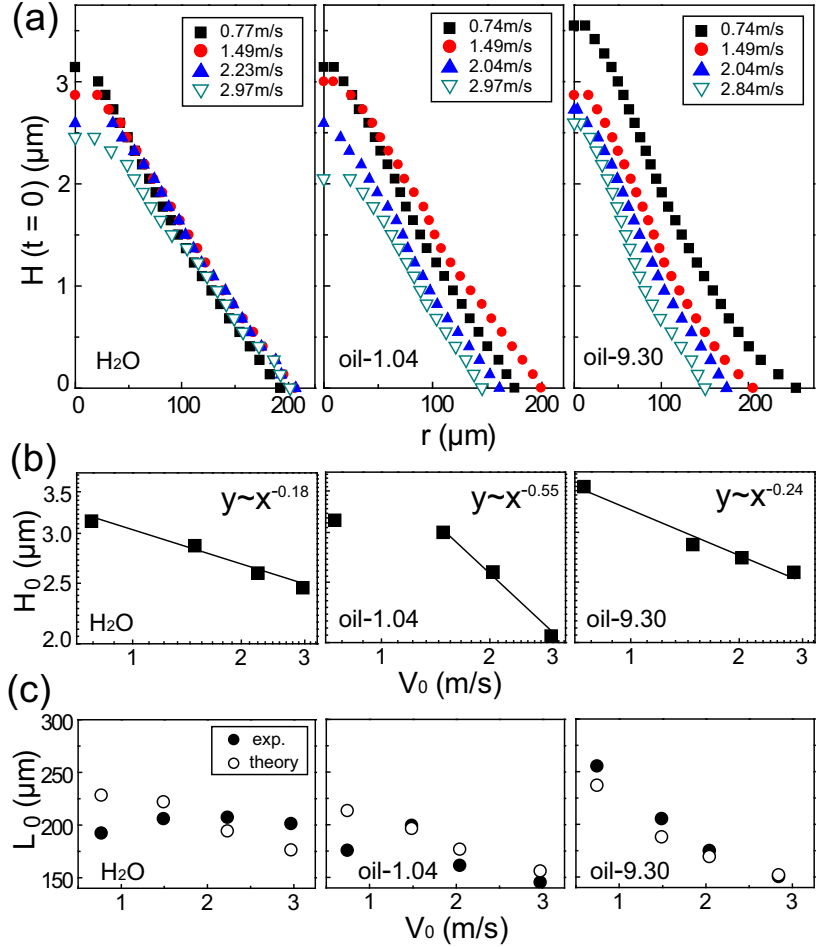}
\caption{(color online) Characterizing the air film profile at
$\mathrm{t=0}$ for different liquids and impact velocities.
(\textit{a}) Air film profiles measured at the moment of contact
($\mathrm{t=0}$) with Newton's rings method. The three panels show
the results for $\mathrm{H_2O}$, oil-1.04 and oil-9.30 respectively.
Different colored curves are measured at different $\mathrm{V_0}$.
(\textit{b}) The maximum film thickness at the center,
$\mathrm{H_0}$, v.s. the impact velocity, $\mathrm{V_0}$. The solid
lines are the best power-law fits within the limited data range.
(\textit{c}) The horizontal radius, $\mathrm{L_0}$, versus impact
velocity, $\mathrm{V_0}$, in different liquids. The solid symbols
are experimental measurements, while the open symbols are
theoretical predictions from an incompressible model,
$\mathrm{L_0=3.8\cdot(4\mu_g/\rho V_0)^{1/3}R^{2/3}}$
(\cite{Peter2012JFM}). Good agreement is observed at high
$\mathrm{V_0}$.}
\end{center}
\end{figure}

\subsection{Evolution of air film after contact ($t\ge 0$)}
The dimple continues to evolve \emph{after} the liquid-solid
contact. The interference patterns for $\mathrm{t\ge0}$ are shown in
Fig.3(a), for an impact with the same condition as in Fig.1(a).
Frame one ($\mathrm{0\mu s}$) shows the Newton's rings from the
initial thin film, with the global contact on a thick black ring at
the edge. During the subsequent contraction ($\mathrm{28-139\mu
s}$), the Newton's rings in the middle remain largely unchanged,
while a grey region without rings grows thicker and thicker at the
edge. This grey region smoothly evolves into an air bubble at the
end of the process (319$\mathrm{\mu s}$ and 2486$\mathrm{\mu s}$).
Such a smooth evolution implies that the thickness of grey region at
$\mathrm{t=319\mu s}$ must be close to the bubble diameter
($\mathrm{d=76\mu m}$) and substantially exceed the coherence length
of light, which explains the lack of Newton's rings. Similarly,
having a thickness larger than the coherence length explains the
absence of rings in the grey region for previous frames as well. To
estimate the thickness there, we assume that the grey region has a
half-circular vertical cross section, as demonstrated by the dashed
curves in Fig.3(b) (the non-circular impression comes from the
different x and y scales).

\begin{figure}
\begin{center}
\includegraphics[width=0.8\textwidth]{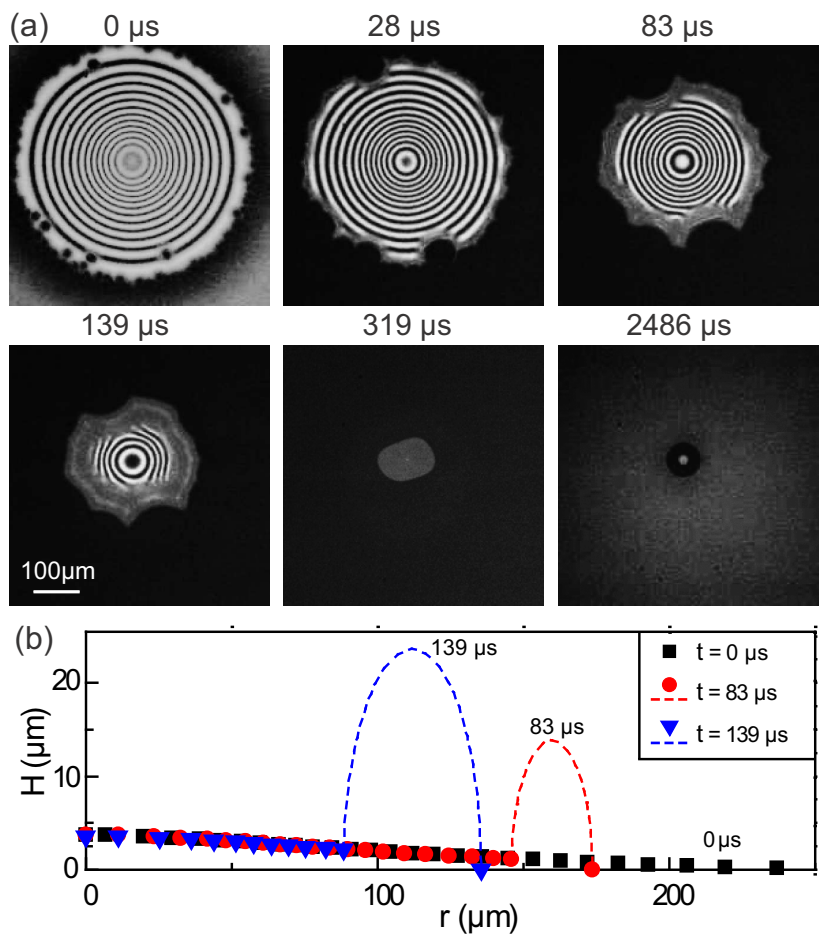}
\caption{(color online) The evolution of trapped air after impact.
The time uncertainty is half of the frame interval, $7\mu s$.
(\textit{a}) The interference patterns for $\mathrm{t\ge0}$ by an
oil-9.30 drop at $\mathrm{V_0=0.74\pm0.01m/s}$. Frame one
($\mathrm{0\mu s}$) shows the Newton's rings from the trapped thin
air film, with the thick black ring at the edge indicating global
liquid-solid contact. During the subsequent contraction
($\mathrm{28-139\mu s}$), the Newton's rings in the middle remain
largely unchanged, while a grey region without rings grows thicker
and thicker at the edge. This grey region smoothly evolves into an
air bubble with diameter $\mathrm{d=76\pm 3\mu m}$ (319$\mathrm{\mu
s}$ and 2486$\mathrm{\mu s}$). (\textit{b}) The thickness profiles
from the center towards the edge along one typical radial direction.
From the patterns in (a), we derive the solid symbols using the
Newton's rings, and estimate the grey region's profile with half
circles as plotted by the dashed curves (Although different scales
in x and y axes produce an non-circular impression). We emphasize
that the dashed curves are from estimate and thus can not be taken
as serious measurements.}
\end{center}
\end{figure}

Combining the thickness measured from the Newton's rings and
estimated at the grey region, we plot the entire thickness profiles
for the $\mathrm{t\ge0}$ frames in Fig.3(b). Different colors
indicate different time $\mathrm{t}$, with the solid symbols from
direct measurement and the dashed curves from the half-circular
estimate. Clearly the air film does not contract with a uniform
thickness; instead the edge grows into a thick rim while the
interior remains thin and flat.

\subsection{The compression at $t=0$}

\begin{figure}
\begin{center}
\includegraphics[width=0.84\textwidth]{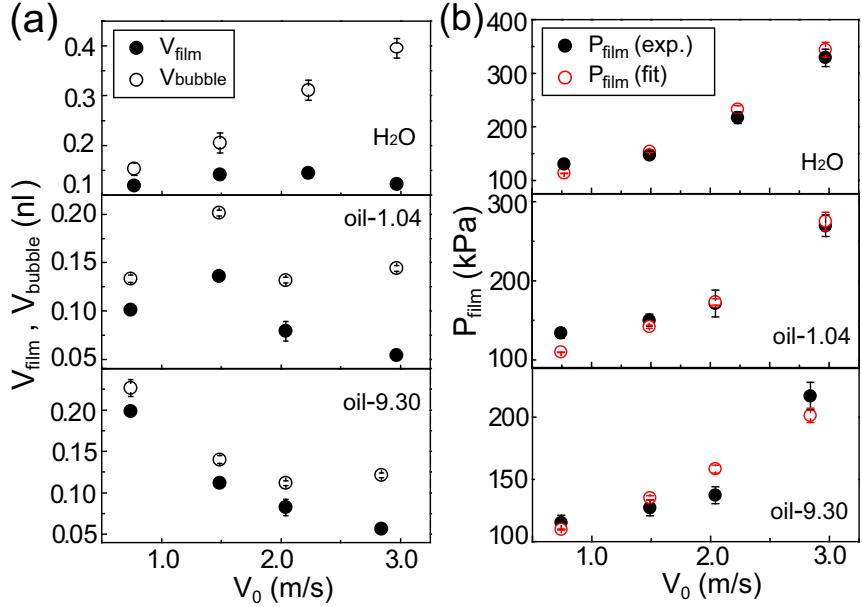}
\caption{(color online) Characterizing the initial pressure (i.e. at
$\mathrm{t=0}$) for different liquids and impact velocities.
(\textit{a}) The volume of the initial air film,
$\mathrm{V_{film}}$, and the final air bubble,
$\mathrm{V_{bubble}}$, v.s. the impact velocity $\mathrm{V_0}$ in
different liquids. Apparently the difference between
$\mathrm{V_{film}}$ and $\mathrm{V_{bubble}}$ increases with
$\mathrm{V_0}$. (\textit{b}) The pressure of air film at
$\mathrm{t=0}$. The solid symbols come from volume measurements
(Eq.\ref{equ:pexp}) and the open symbols are calculated from the
expression (Eq.\ref{equ:pmodel}). They agree reasonably well. The
fitting parameter, C, has the values: $\mathrm{C = 0.33}$
($\mathrm{H_2O}$), 0.33 (oil-1.04) and 0.23 (oil-9.30). It does not
change with surface tension but decreases with viscosity.}
\end{center}
\end{figure}

Is the trapped air initially ($\mathrm{t=0}$) compressed or not? We
address this fundamental question by accurately measuring and
comparing the volumes at the initial ($\mathrm{t=0}$) and the final
states. The volume at $\mathrm{t=0}$ is directly obtained from the
thickness profile H(r); and the volume of the final air bubble is
accurately determined from its diameter. To make sure that the
bubble is a perfect sphere instead of a cap intersected by the
substrate, we keep track of the bubble until it leaves the surface
and rises upward. Since the bubble is a perfect sphere after leaving
the surface, we can unambiguously determine its volume from the
diameter. The exact values of $\mathrm{V_{film}}$ and
$\mathrm{V_{bubble}}$ are shown in Fig.4(a) for different velocities
and liquids. The plot reveals a considerable volume increase from
film to bubble, proving that the trapped air at $\mathrm{t=0}$ is
indeed \emph{compressed}.

To further estimate the pressure at $\mathrm{t=0}$, we need to
clarify whether the compression is isothermal or adiabatic. It
depends on the time scale of air entrapment, $\tau$, compared with
the time scale during which a thermal equilibrium can be reached,
$\tau'$. In our experiment, $\tau\sim \mathrm{H_0/V_0}\sim1\mu s$;
while $\tau'$ is determined by the rate of thermal conduction. Our
system is a thin air film with the upper and lower boundaries
(liquid drop and glass substrate respectively) at the room
temperature. Suppose there is a temperature change $\mathrm{\Delta
T}$ in the middle (from gas compression or viscous heating or other
reasons), the thermal gradient then becomes $\mathrm{\Delta
T/(H_0/2)}$ and the heat flux per unit time is $\mathrm{dQ/dt\sim
2\cdot k\cdot A\cdot \Delta T/(H_0/2)}$, with $\mathrm{k}$ being the
thermal conductivity of air and $\mathrm{A}$ being the horizontal
cross section area; the factor of 2 comes from the existence of two
boundaries. This heat flux will bring the system back to equilibrium
during the time scale: $\mathrm{\tau'\sim c_p\cdot\rho_{air}\cdot
v_{air}\cdot \Delta T/(dQ/dt)\sim c_p\cdot \rho_{air}\cdot
H_0^2/12k}$, with $\mathrm{c_p}$ being the air's specific heat per
unit mass and $\mathrm{v_{air}\sim\frac{1}{3}A\cdot H_0}$ being the
volume of trapped air. Plugging in the values for atmospheric
pressure and room temperature, $\mathrm{c_p = 10^3 J / (kg\cdot K),
\rho_{air} = 1.2 kg/m^{3}, k = 0.026 J / (s\cdot m\cdot K)}$, we
obtain: $\mathrm{\tau'}\sim0.03\mu s\ll\tau\sim1\mu s$. Therefore,
the trapped air reaches thermal equilibrium rather rapidly, due to
its thin thickness and small heat capacity, and the compression in
our experiment can be considered as \emph{isothermal}. Subsequently,
the compressed air film expands to an uncompressed air bubble, again
in an isothermal manner since expansion is even slower than
compression. Thus we can obtain the initial air film pressure,
$\mathrm{P_{film}}$, from the isothermal equation of state:

\begin{equation}
\mathrm{P_{film}=P_{bubble}\cdot\frac{V_{bubble}}{V_{film}}=P_0\cdot\frac{V_{bubble}}{V_{film}}}
\label{equ:pexp}
\end{equation}

\noindent Here $\mathrm{P_{bubble}=P_0=101kPa}$ since the bubble is
uncompressed (the curved surface of bubble adds a negligibly small
Laplace pressure around $1$kPa to $\mathrm{P_{bubble}}$). The exact
values of $\mathrm{P_{film}}$ are shown as solid symbols in
Fig.4(b). Apparently, $\mathrm{P_{film}}$ increases dramatically
with impact speed, varying from very close to $\mathrm{P_0}$ to
several $\mathrm{P_0}$.

This velocity dependence can be qualitatively understood by a force
balance estimate. Immediately before the contact, the compressed air
in the dimple locally decelerates the liquid to an almost complete
stop in vertical direction (see Fig.1(b) main panel), on the length
scale of the dimple's radius $\mathrm{L_0}$ (see Fig.1(b) inset).
The deceleration occurs during the time scale $\mathrm{\tau \sim
H_0/V_0}$, with the magnitude $\mathrm{a\sim V_0/\tau\sim
V_0^2/H_0}$. Since the affected liquid has the mass
$\mathrm{m}\sim\rho\mathrm{L_0^3}$, we get the force:
$\mathrm{f=m\cdot a\sim C\rho L_0^3 V_0^2/H_0}$, with C being a
pre-factor of order unity. This force naturally comes from the
excess pressure of compressed air multiplying the area:
$\mathrm{f\sim(P_{film}-P_0)\cdot L_0^2}$. Making the two force
expressions equal (\cite{Brenner2009PRL, Brenner2010JFM}) and
solving for $\mathrm{P_{film}}$ yield:

\begin{equation}
\mathrm{P_{film} =  P_0 + C} \cdot\rho\mathrm{V_0^{2} \cdot
\frac{L_0}{H_0}}
\label{equ:pmodel}
\end{equation}

This expression has the same trend as the measurements: at small
$\mathrm{V_0}$, $\mathrm{P_{film}}$ approaches $\mathrm{P_{0}}$;
while it rises rapidly with $\mathrm{V_0}$ due to the large
geometric factor, $\mathrm{L_0/H_0\sim100\gg1}$. The values
calculated from this expression are plotted as open symbols in
Fig.4(b), and agree reasonably well with the measurements (solid
symbols), for different liquids at various impact speeds. The only
fitting parameter, C, is indeed of order unity: C = 0.33
($\mathrm{H_2O}$), 0.33 (oil-1.04) and 0.23 (oil-9.30). Clearly C
does not change with surface tension but decreases with viscosity,
suggesting that under the same impact condition the air trapped in a
more viscous drop is less compressed.

Comparing with the existing compressible model
(\cite{Brenner2009PRL, Brenner2010JFM}), our experimental range
overlaps with their transition region from ``incompressible regime''
to ``compressible regime'' (i.e., $\mathrm{0.1<\epsilon^{-1}<10}$).
Indeed the data in Fig.4(b) verify such a predicted transition: at
low $\mathrm{V_0}$ the pressure is close to $\mathrm{P_0}$ and the
compression is small; while significant compression appears above
$\mathrm{V_0\sim1.5m/s}$ (corresponding to $\mathrm{\epsilon\sim1}$
in the theory). We also find that the initial impact pressure,
$\mathrm{P_{film}}$, exceeding the liquid inertia,
$\rho\mathrm{V_0^{2}}$, by a large factor of
$\mathrm{L_0/H_0\sim100}$. This comes from the dramatic deceleration
of liquid under the thin gap geometry.

\subsection{The contraction at $t>0$}
From compressed air film to uncompressed air bubble, the detailed
process requires further clarification. We measure this process from
interference patterns and demonstrate the evolution in Fig.5(a). At
the moment of contact (t = 0), the trapped air is thin and flat with
a pressure higher than $\mathrm{P_0}$. After the contact (t $>$ 0),
however, the air pressure drops sharply as the impact pressure
decreases, resulting in a rapid volume expansion at the edge of
film. This non-uniform expansion may come from the non-uniform
pressure distribution after contact: the center of the film is right
below the stagnation point of liquid and thus experiences higher
impact pressure; while the edge of air film experiences less impact
pressure and more easily to expand. This expansion produces a rim
much thicker than the interior, as illustrated in Fig.5(a). We
estimate the rim's volume by assuming a half-circular profile and
deduce the pressure from volume estimate. We find that the pressure
rapidly drops to uncompressed value within a few tens of
microseconds. The uncompressed thick-rimmed structure subsequently
contracts into an air bubble.

\begin{figure}
\begin{center}
\includegraphics[width=0.8\textwidth]{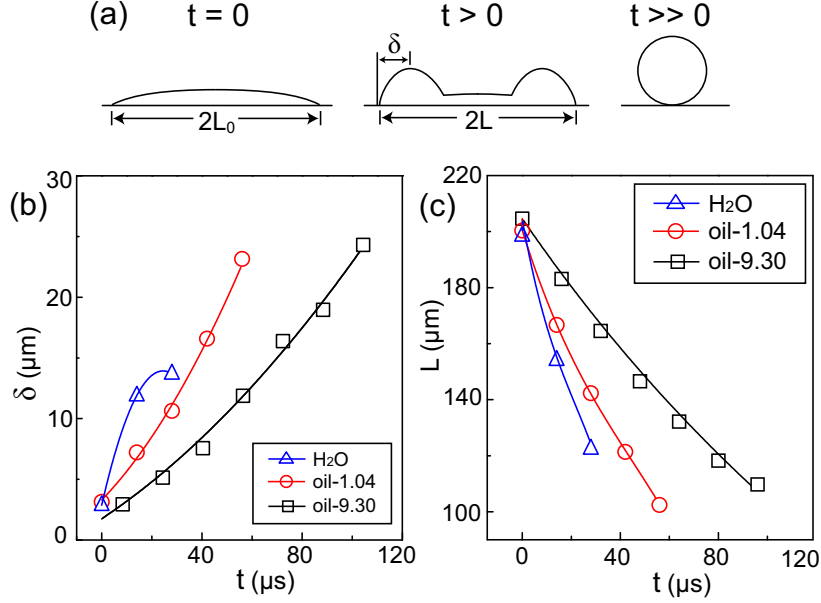}
\caption{(color online) Evolution of trapped air during contraction.
(\textit{a}) Schematics illustrating the vertical profile evolution
during contraction. The air film turns into uncompressed
thick-rimmed structure within a few tens of microseconds, and then
contracts into an air bubble. (\textit{b}) Measuring $\delta(t)$
experimentally. The three curves come from the three liquids at the
same impact velocity, $\mathrm{V_0=1.49\pm0.02m/s}$. The open
symbols are our measurements and the solid curves are
high-order-polynomial fits. These functions are plugged into Eq.3.4
to calculate $\mathrm{L(t)}$. (\textit{c}) The comparison of
$\mathrm{L(t)}$ between the measurements and the calculation. The
open symbols come from direct measurements while the curves are
calculated from Eq.3.4. With the same fitting parameters,
$\mathrm{C_1=0.63}$ and $\mathrm{C_2=1.54}$, we obtain excellent
agreement for different liquids. The air film contracts the fastest
in $\mathrm{H_2O}$, due to the large surface tension and low
viscosity. It contracts slower in oil-1.04, because of the reduced
surface tension. The motion is the slowest in oil-9.30, due to the
effects of both low surface tension and high viscosity.}
\end{center}
\end{figure}

Therefore, the contraction process is mostly under uncompressed
condition except at the very beginning, consistent with the
incompressible assumption in previous contraction measurements
\citep{Thoroddsen2003JFM, Thoroddsen2005JFM}. However, the previous
study assumes a uniform film thickness, while we believe that the
thick rim plays an important role. We take the characteristic size
of the rim, $\mathrm{\delta}$ (see Fig.5(a)), as the dominant length
scale during contraction: the surface tension provides the driving
stress, $\mathrm{\sigma/\delta}$, which is balanced by the liquid
inertial, $\mathrm{\rho v^2}$, and the viscous stress, $\mathrm{\mu
v/\delta}$. We note that the buoyant effect may also drive the
dewetting of air film. However when compared with the surface
tension, the buoyant effect is completely negligible during the
contraction period studied here: $\Delta \rho \cdot V \cdot g/A \sim
0.1Pa \ll \sigma/\delta\sim 10^3Pa$. Therefore we neglect the
buoyant effect and use the surface tension as the solely driving
force:

\begin{equation}
\mathrm{\frac{\sigma}{\delta(t)}=C_1\cdot\rho v(t)^{2} +
C_2\cdot\mu\frac{v(t)}{\delta(t)}}
\label{equ:cont}
\end{equation}

Here $\mathrm{v(t)}$ is the instantaneous contraction velocity at
the edge, and $\mathrm{C_1}$, $\mathrm{C_2}$ are pre-factors of
order one. This expression relates the contraction speed
$\mathrm{v(t)}$ to the rim size $\mathrm{\delta(t)}$. From
$\mathrm{v(t)}$, we can further calculate the pattern's radius,
$\mathrm{L(t)}$ (see Fig.5(a)), via the relationship
$\mathrm{v(t)=-dL(t)/dt}$. Thus we can obtain L(t) by solving for
$\mathrm{v(t)}$ in Eq.\ref{equ:cont} and then integrate it:

\begin{eqnarray}
\mathrm{L(t)} &=& \mathrm{L_0 - \int_{0}^{t} v(\tau)d\tau} \\
&=& \mathrm{L_0- \int_{0}^{t} \frac{-C_2\mu +
\sqrt{C_2^2\mu^{2}+4C_1\rho\sigma\delta(\tau)}}{2C_1\rho\delta(\tau)}
d\tau}\nonumber
\end{eqnarray}

Therefore, $\mathrm{L(t)}$ can be calculated from the knowledge of
$\mathrm{\delta(\tau)}$. We obtain the function
$\mathrm{\delta(\tau)}$ by first measuring $\delta$ experimentally
and then fitting the data with high order polynomials, as shown in
Fig.5(b). On the other hand, $\mathrm{L(t)}$ can be independently
measured from our images. To directly measure L(t), we make the best
circular fit for the edge of the pattern, and obtain L(t) from the
radius of the circle. The results of two approaches are compared in
Fig.5(c) for different liquids. The calculated curves match the
measured symbols quite well, confirming the validity of the
expression in Eq.\ref{equ:cont}. In addition, the two fitting
parameters, $\mathrm{C_1=0.63}$ and $\mathrm{C_2=1.54}$, are
universally valid for all liquids, and are indeed of order unity.
The air film contracts the fastest in $\mathrm{H_2O}$, due to the
large surface tension and low viscosity. It contracts slower in
oil-1.04, because of the reduced surface tension. The motion is the
slowest in oil-9.30, due to the effects of both low surface tension
and high viscosity. In conclusion, the contraction is determined by
the complex interaction between surface tension, inertia, and
viscous effects, as described by Eq.\ref{equ:cont}. Because the
dynamic contact angle during contraction is different from the
static value and can not be measured directly, its role for air film
contraction is still unclear.

\section{Conclusions}\label{sec:con}
In this work, we couple high-speed photography with optical
interference to study the formation and evolution of air entrapment
in liquid-solid impacts. We find a compressed air film formed before
liquid touches solid, with an internal pressure significantly higher
than the atmospheric value. After contact, the air film vertically
expands at the edge, reducing the pressure within a few tens of
microseconds and making a rim much thicker than the interior. This
thick-rimmed structure subsequently contracts into an air bubble,
driven by the complex interaction between surface tension, inertia,
and viscous effects. Our investigation provides explicit information
for the initial impact pressure as well as the detailed profile
transformation during the subsequent evolution. This knowledge may
benefit impact related applications such as surface coating and
splash control.

This project is supported by the Hong Kong RGC CUHK404211,
CUHK404912 and direct grant 2060418.

\bibliographystyle{jfm}

\bibliography{jfm-instructions}

\end{document}